\documentclass[pdftex,twocolumn,epjc3]{svjour3}

\usepackage[T1]{fontenc}
\usepackage[utf8]{inputenc}
\usepackage{graphicx}
\usepackage{xcolor}
\usepackage{mathtools}
\usepackage{amssymb}
\usepackage{lmodern}
\usepackage{bm}
\usepackage{siunitx}
\usepackage[final]{microtype}
\usepackage{hyperref}
\usepackage{multirow}

\allowdisplaybreaks 

\journalname{EPJST}

\newcommand{\mytitle}{Quark mass dependence of a QCD critical point and structure of the Columbia plot}

\newcommand{\JLU}{%
    Institut f\"{u}r Theoretische Physik,
    Justus-Liebig-Universit\"{a}t Gie\ss{}en,
    35392 Gie\ss{}en,
    Germany
}

\newcommand{\HFHF}{%
    Helmholtz Forschungsakademie Hessen f\"{u}r FAIR (HFHF),
    GSI Helmholtzzentrum f\"{u}r Schwerionenforschung,
    Campus Gie\ss{}en,
    35392 Gie\ss{}en,
    Germany
}

\hypersetup{%
    pdftitle = {\mytitle},
    pdfauthor = {Julian Bernhardt, Christian S. Fischer},
    bookmarksopen = true,
    colorlinks = true,
    linkcolor = green!40!black,
    citecolor = blue!50!black,
    urlcolor = blue!50!black
}

\graphicspath{{./Graphics/}}

\DeclareSIUnit{\MeV}{\mega\electronvolt}
\DeclareSIUnit{\GeV}{\giga\electronvolt}
\DeclareSIUnit{\fm}{\femto\meter}

\DeclareMathOperator{\Tr}{Tr}

\DeclarePairedDelimiter{\expval}{\langle}{\rangle}

\renewcommand*{\vec}[1]{\bm{#1}}
\newcommand*{\+}{\hspace*{.08335em}}

\newcommand*{\dd}{\mathrm{d}}
\newcommand*{\ii}{\mathrm{i}}

\newcommand*{\bbZ}{\mathbb{Z}}

\newcommand*{\muCEP}{\mu_{\textup{B}}^{\textup{CEP}}}
\newcommand*{\TCEP}{T_{\textup{CEP}}}
\newcommand*{\Tc}{T_{\textup{c}}}
\newcommand*{\Nc}{N_{c}}
\newcommand*{\Nf}{N_{f}}

\newcommand*{\OO}{\textup{O}}
\newcommand*{\gs}{g}

\newcommand*{\ups}{\textup{s}}
\newcommand*{\massu}{m_{\textup{u}}}
\newcommand*{\massd}{m_{\textup{d}}}

\newcommand*{\massuds}{m_{\textup{u,d,s}}}
\newcommand*{\massl}{m_{\ell}}
\newcommand*{\masss}{m_{\textup{s}}}

\newcommand*{\mus}{\mu_{\textup{s}}}
\newcommand*{\muB}{\mu_{\textup{B}}}

\newcommand*{\SU}{\textup{SU}}
\newcommand*{\SUA}{\textup{SU}_{A}}

\newcommand*{\UA}{\textup{U}_{A}}

\newcommand*{\mutri}{\mu_{\textup{B}}^{\textup{tri}}}
\newcommand*{\Ttri}{T_{\textup{tri}}}

\definecolor{dgreen}{rgb}{0.1,0.5,0.1}
\definecolor{lblue}{rgb}{0.2,0.35,1}
\definecolor{webred}{rgb}{0.75,0,0}

\begin{document}

\title{\mytitle}

\author{Julian Bernhardt\thanksref{addr1} \and Christian S. Fischer\thanksref{e2,addr1,addr2}}

\thankstext{e2}{e-mail: christian.fischer@theo.physik.uni-giessen.de}

\institute{\JLU \label{addr1}	\and	\HFHF \label{addr2}	}

\maketitle

\begin{abstract}
We study the quark-mass dependence of the QCD critical point at varying bare
up/down quark masses with fixed strange quark mass. We explore the corresponding
second-order critical surface in the three-dimensional Columbia plot and study
the extension of the associated crossover hyperplane at real and imaginary
baryon chemical potential. To this end, we employ a by now well-tested
combination of lattice Yang--Mills theory and a (truncated) version of
Dyson--Schwinger equations at $\Nf=2+1$ quark flavours. We find evidence for a
positive curvature of the second order surface at (large) real chemical
potential and a crossover region for imaginary chemical potential. Our results
support the notion of a tricritical point at finite chemical potential in the
chiral limit.
\end{abstract}

\section{\label{sec:introduction}%
    Introduction
}

In the past years, results from lattice QCD led to a wide-spread consensus on
the crossover nature of QCD's chiral transition at zero chemical potential
\cite{Aoki:2006we,Borsanyi:2010bp,Bazavov:2011nk,HotQCD:2018pds,Borsanyi:2020fev}.
 The pseudocritical temperature, varying only slightly between different 
definitions of the chiral order parameter, has been localized around $\Tc 
\approx \SIrange{155}{160}{\MeV}$ and thermodynamic properties of hot matter at 
zero and small to medium chemical potential have been determined with ever 
increasing precision 
\cite{Borsanyi:2010cj,Borsanyi:2013bia,HotQCD:2014kol,Ding:2015ona,Bazavov:2017dus,Bollweg:2022fqq,Borsanyi:2025dyp}. 
There is also 
growing consensus that the crossover region extends far into the real chemical 
potential region. No QCD endpoint is found in the region of the 
temperature--baryon-chemical-potential plane $(T,\muB)$ with $\muB / T < 2.5$ 
\cite{HotQCD:2018pds,Borsanyi:2020fev} and very recent results suggest this 
crossover area stretches at least up to chemical potentials of $\muB < 450$ MeV 
\cite{Borsanyi:2025dyp}.

Results from functional approaches to QCD, i.e., via Dyson--Schwinger equations
(DSE) and/or the functional renormalization group (FRG), with realistic
truncations of the Yang-Mills sector of QCD confirm these findings
\cite{Fischer:2014ata,Isserstedt:2019pgx,Fu:2019hdw,Gao:2020qsj,Gao:2020fbl,Gunkel:2021oya,Lu:2023mkn} 
and argue to exclude a CEP in the region
$(T,\muB)$ with $\muB / T < 4$. The three works with arguably most advanced
truncations find a critical endpoint in a region around $(\TCEP,\muCEP) =
(110,620) \, \si{\MeV}$ \cite{Fu:2019hdw,Gao:2020fbl,Gunkel:2021oya}, albeit
with sizeable systematic error\footnote{For an estimate of the lower bound of
such an error one may quote twice the spread of the three results. This amounts
to $(\Delta \TCEP,\Delta \muCEP) = (10,40)\, \si{\MeV}$.}. Results from
holographic QCD \cite{Hippert:2023bel} and extrapolations of the positions of
Lee--Yang zeros \cite{Basar:2023nkp,Clarke:2024ugt} as well as thermodynamic
quantities from lattice QCD \cite{Shah:2024img} seem to confirm the notion of a
critical point at large chemical potential.

\begin{figure}
    \centering
    \includegraphics[scale=0.48]{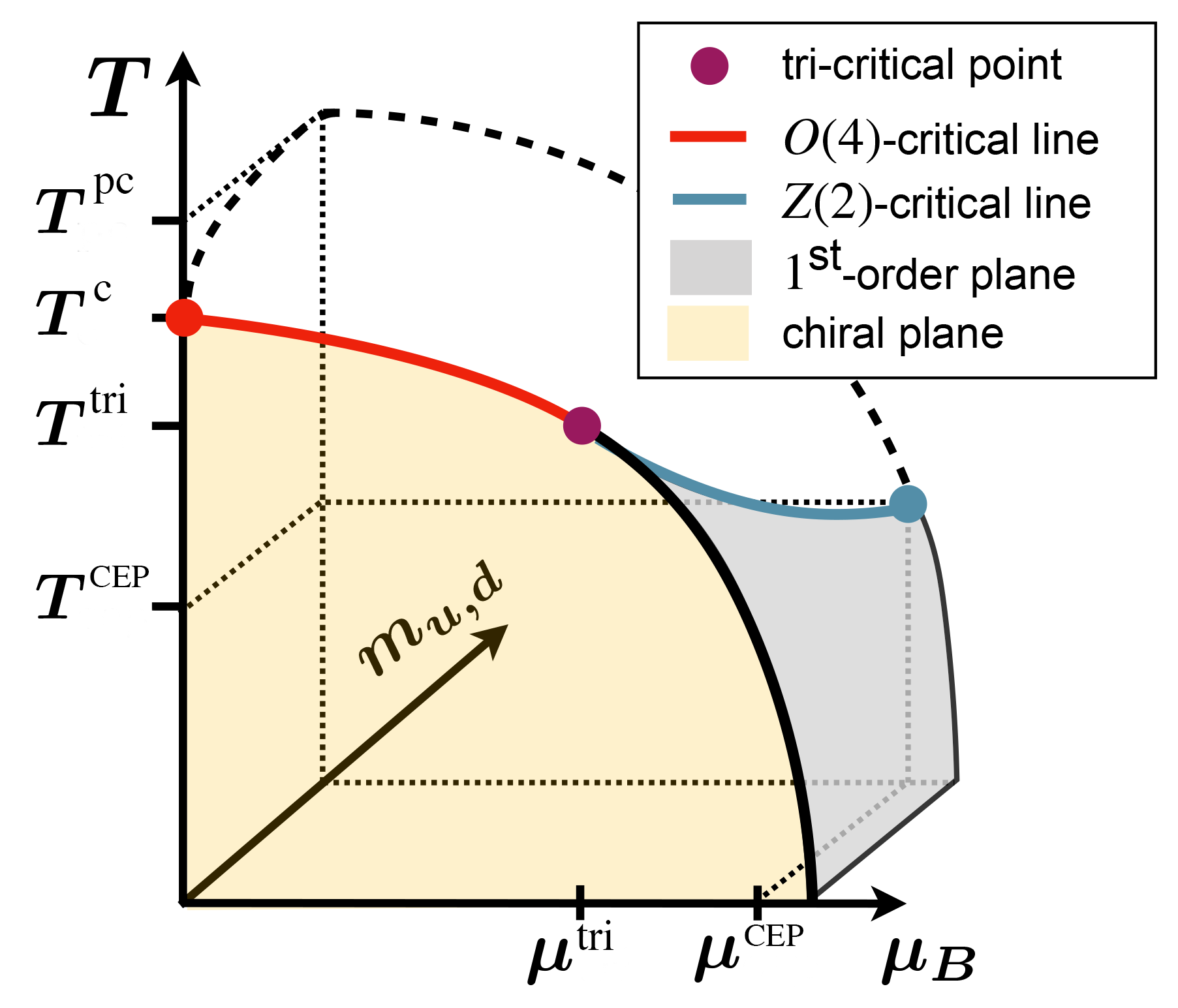}
    \caption{\label{fig:phases}%
        Sketch of the QCD phase diagram in temperature $T$ and baryon chemical
        potential $\muB$ for varying degenerate light quark masses
        $\massl=\massu=\massd$. Adapted from Ref.~\cite{Ding:2024sux}.
    }
\end{figure}

\begin{figure*}
    \centering
    \includegraphics[scale=1.3]{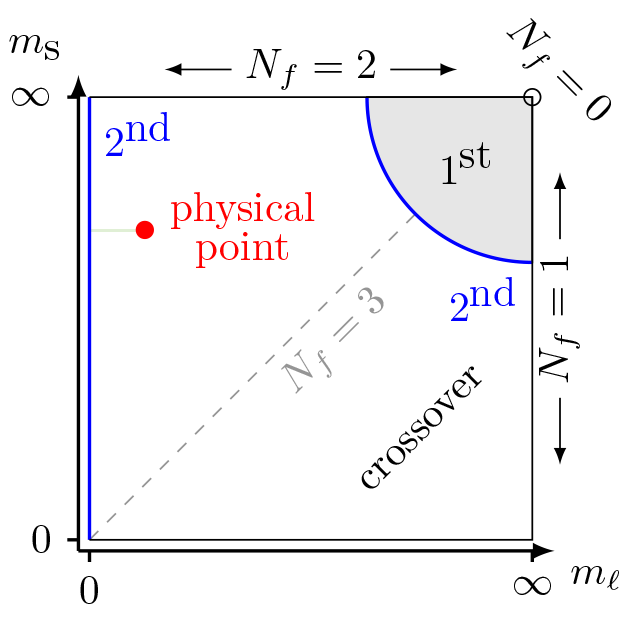}
    \hfil
    \includegraphics[scale=0.5]{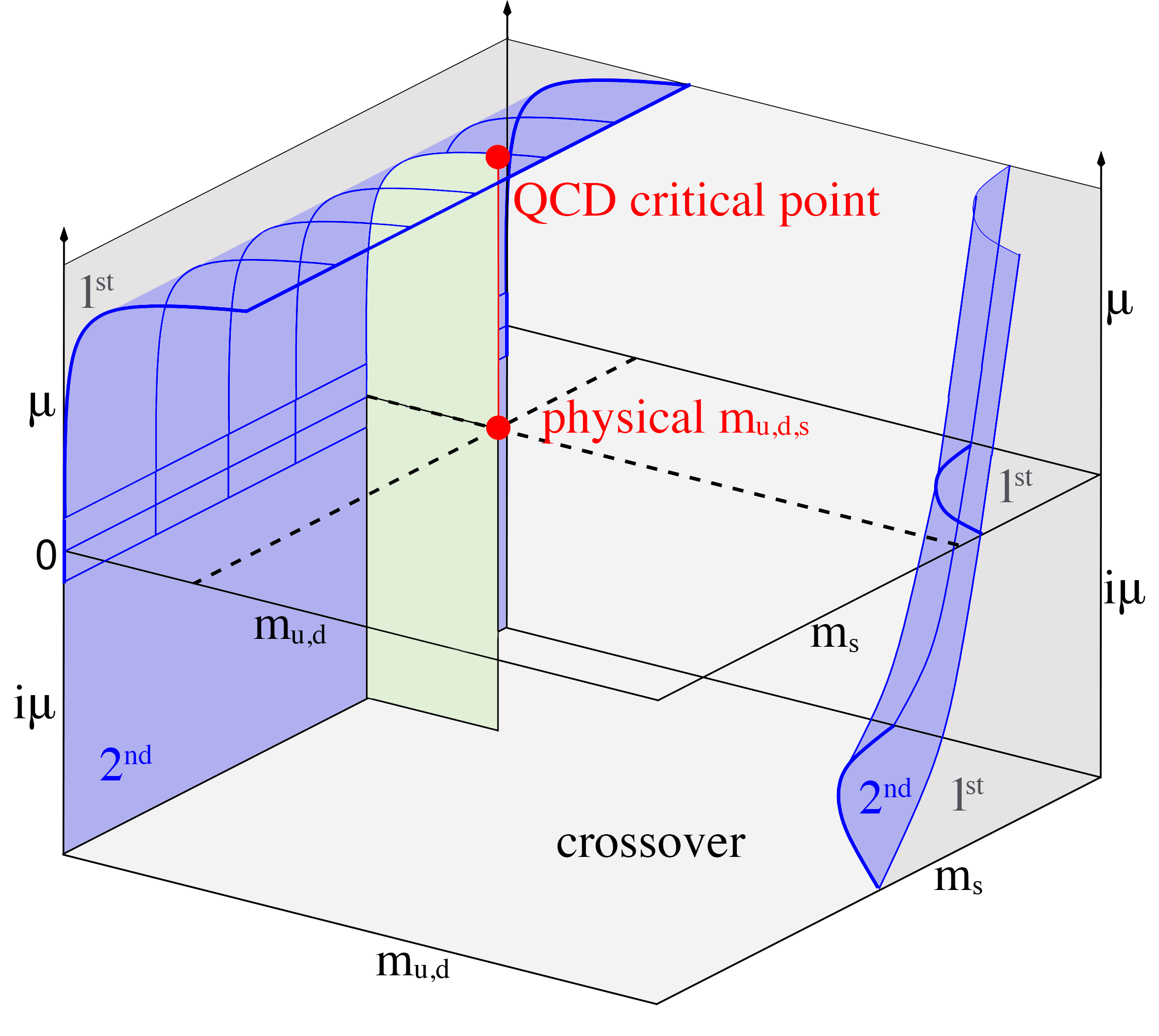}
    \caption{\label{fig:columbia-plot}%
        Left: Columbia plot \cite{Brown:1990ev} of phase-transition orders at
        nonzero temperature and vanishing chemical potential as functions of
        quark masses $\massl = \massu = \massd$ and $\masss$. Whether the lower
        left corner, $\massuds \rightarrow 0$, is second or first order is
        currently under debate.
        Right: 3$d$-version of the Columbia plot with imaginary and real
        chemical potential as additional axis. The light green area is explored
        in this work.
        See text for further explanations.
    }
\end{figure*}

For systematic reasons, it is very interesting to connect this critical point at
physical quark masses with the behaviour of QCD at unphysical quark masses such
as, e.g., the chiral limit of vanishing light (up/down) quark masses. This is
the topic of this work. In Fig.~\ref{fig:phases}, we display a sketch of the
phase diagram of (2+1)-flavour QCD (adapted from Ref.~\cite{Ding:2024sux})
visualizing the conjectured phase structure in the temperature and
chemical-potential plane as a function of light quark mass. Consider first the
point $T^c$ at vanishing chemical potential and vanishing light-quark masses. As
indicated in the plot, its universality class is expected to be the one of 3$d$,
$\OO(4)$ spin models \cite{Pisarski:1983ms}, although depending on the fate of
the axial $\UA(1)$ anomaly (see below), a larger symmetry group cannot be ruled
out at present. In the 2$d$-Columbia plot on the left side of
Fig.~\ref{fig:columbia-plot}, this point is approached by starting from the
physical point and then going horizontally along the green line to the left into
the chiral limit $\massl=\massu=\massd=0$. Strong indications for the
second-order nature of this point have been reported from lattice QCD
\cite{HotQCD:2019xnw,Cuteri:2021ikv} and functional methods
\cite{Braun:2020ada,Gao:2021vsf,Bernhardt:2023hpr}.

In the $\muB=0$-plane, this point is connected to the two- and three-flavour
massless corners of the Columbia plot by a line with varying strange-quark
masses from zero to infinity. This line is governed by the chiral transition and
the corresponding axial symmetries $\UA(1) \times \SUA(\Nf)$. Whereas the latter
one is broken dynamically at low temperatures (and always explicitly by nonzero
quark masses), the former one is broken anomalously. Both the dynamical and
anomalous breaking can be restored at large temperatures, albeit the
corresponding transition temperatures may differ from each other. Whether
$\UA(1)$ remains broken at the chiral $\SUA(\Nf)$ transition is an open question
with conflicting indications \cite{Brandt:2016daq,Tomiya:2016jwr,Aoki:2021qws,HotQCD:2012vvd,Buchoff:2013nra,Bhattacharya:2014ara,Dick:2015twa,Ding:2020xlj,Kaczmarek:2021ser}. 
The fate of the $\UA(1)$ symmetry is expected
to affect the order of the chiral $\SUA(\Nf)$ transition, see e.g.
\cite{Pisarski:1983ms,Resch:2017vjs,Pisarski:2024esv,Giacosa:2024orp} and Refs.
therein. In the past years, the possible scenario of a second-order transition
for all nonzero masses of the strange quark emerged and even the long-standing
notion of the limit $\massuds \rightarrow 0$ as being first order has been
questioned \cite{deForcrand:2017cgb}. Recent lattice results indeed indicate a
second-order nature of this point \cite{Cuteri:2021ikv,Dini:2021hug}. How such a
scenario can be accommodated and explained by effective models is currently
under discussion \cite{Resch:2017vjs,Fejos:2022mso,Pisarski:2024esv,Giacosa:2024orp}.

The second-order scenario is also supported by functional methods
\cite{Bernhardt:2023hpr} and we therefore chose to display it in
Fig.~\ref{fig:columbia-plot}. It seems to persist for small real and imaginary
chemical potential at least to values of $\muB^2 = \pm (\SI{30}{\MeV})^2$, as
indicated in the 3$d$-Columbia plot on the right hand side of
Fig.~\ref{fig:columbia-plot} by the two blue lines parallel to the chiral
$\muB=0$-line. The purpose of this work is to explore the fate of the chiral
transition in the large real and imaginary chemical potential region at fixed,
physical strange quark mass, but varying $\massl$ from the physical point to the
chiral limit. In the 3$d$-Columbia plot on the right hand side of
Fig.~\ref{fig:columbia-plot} this corresponds to the area shaded in light green.
We will explore the green area at imaginary chemical potential and verify that
the chiral transition is a crossover for $0 < \massl \le \massl^{phys}$, i.e.
that the critical second-order surface does not extend into the finite
light-quark mass region. Indications for this behaviour have been seen
previously on the lattice \cite{DAmbrosio:2022kig,Cuteri:2022vwk}. For real
chemical potential, we seek to explore the fate of the QCD critical point when
the light-quark masses are lowered towards the chiral limit. In
Fig.~\ref{fig:phases}, this corresponds to the blue line connecting the critical
endpoint at $(\TCEP,\muCEP)$ to the tricritical point $(\Ttri,\mutri)$ in the
chiral-limit plane. The curvature of this line is governed by thermodynamics.
Model calculations confirm that the transition temperature of the tricritical
point is larger than the one of the critical endpoint \cite{Halasz:1998qr} and
therefore both the temperatures $\Tc$ and $\Ttri$ are expected to give upper
bounds for $\TCEP$ \cite{Karsch:2019mbv}. In this work, we will further
scrutinize this notion by following the blue line from the CEP towards the
chiral limit of vanishing light-quark masses (while keeping $\masss$ fixed). In
the 3$d$-Columbia plot of Fig.~\ref{fig:columbia-plot}, this corresponds to
tracing the second-order critical surface from the QCD critical point to the
left, i.e., on the upper edge of the light-green-shaded area.

The paper is organized as follows. In the next Section~\ref{sec:framework}, we
briefly summarize the framework of Dyson--Schwinger equations that has been used
previously to study the location of the critical endpoint \cite{Gunkel:2021oya}
and the light quark chiral limit of the 3$d$-Columbia plot
\cite{Bernhardt:2023hpr}. We discuss our results in Section~\ref{sec:results}
and conclude in Section~\ref{sec:summary}.

\section{\label{sec:framework}%
    Framework
}

\subsection{\label{subsec:dse}%
    Dyson--Schwinger equations
}

The central quantity to study the Columbia plot is the light-quark condensate as
an order parameter for chiral symmetry breaking. It is extracted from the quark
propagator $S_f(q)$ via the relation
\begin{equation}\label{eq:condensate}
    \expval{\bar{\psi} \psi}_{f}
    =
    -3 Z_{2}^{f} Z_{m_{f}} T
    \sum_{\omega_{q}} \int \frac{\dd^{3} q}{(2 \pi)^{3}}
    \Tr\bigl[ S_f(q) \bigr]
    \,.
\end{equation}
Here, the factor three stems from the colour trace ($\Nc = 3$) and the quark
flavour is denoted by $f \in \{\ell, \ups\}$ with $\ell \in \{\textup{u,d}\}$.
The explicit form of $S_f(q)$ will be discussed below. The quark condensate is
quadratically divergent for all flavours with a nonzero bare-quark mass because
of a contribution proportional to $m_{f} \Lambda^{2}$ where $\Lambda$ denotes a
regularization scale at large momenta. A suitably regularized version of the
condensate is therefore given by the difference
\begin{equation}\label{eq:condensate_subtracted}
    \Delta_{\ell\ups}
    =
    \expval{\bar{\psi} \psi}_{\ell}
    -
    \frac{Z_{m}^{\ell}}{Z_{m}^{\ups}}
    \frac{m_{\ell}}{m_{\ups}} \+
    \expval{\bar{\psi} \psi}_{\ups}
    \,.
\end{equation}
Note that we are working with renormalized quantities, hence the appearance of
the mass renormalization constants $Z_{m}^{f}$ in order to preserve
multiplicative renormalizability. In the case of massless light quarks, the
subtracted condensate reduces to the unsubtracted one, which is also finite in
this case. The chiral susceptibility is defined as the derivative of the
regularized condensate with respect to the light-quark mass:
\begin{equation}
    \chi_{\ell\ups}^{m}
    =
    \frac{\partial}{\partial \massl}
    \Delta_{\ell\ups}
    \,.
\end{equation}
We use this quantity to monitor the order of the chiral transition. Up to
normalization factors and for $\massl = \massu = \massd$, this definition is
equivalent to the ones used in Refs.~\cite{HotQCD:2019xnw,Braun:2020ada}.

The inverse, dressed (i.e., full) quark propagator $S_{f}$ at nonzero
temperature $T$ and quark chemical potential $\mu_{f}$ is given by
\begin{equation}\label{eq:quark_propagator}
    S_{f}^{-1}(p)
    =
    \ii \gamma_{4} \tilde{\omega}_{n}^{f} C_{f}(p)
    +
    \ii \vec{\gamma} \cdot \vec{p} A_{f}(p)
    +
    B_{f}(p)
    \,.
\end{equation}
Here, $p = (\vec{p}, \tilde{\omega}_{n})$ represents the four-momentum, while
$\tilde{\omega}_{n}^{f} = \omega_{n} + \ii \mu_{f}$ denotes a combination of the
fermionic Matsubara frequencies $\omega_{n} = (2n + 1) \pi T$, $n \in \bbZ$,
with the chemical potential. All non-perturbative information such as the
non-trivial momentum dependence is carried by the quark dressing functions
$A_{f}$, $B_{f}$ and $C_{f}$.

\begin{figure}
    \centering
    \includegraphics[width=\columnwidth]{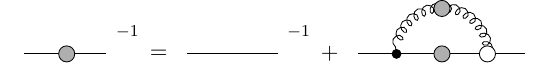}
    \\[0.25em]
    \includegraphics[width=\columnwidth]{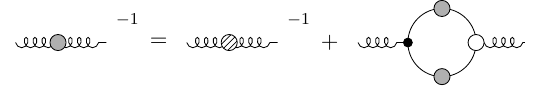}
    \caption{\label{fig:dses}%
        General form of the DSE for the quark propagator (top) and truncated
        gluon DSE (bottom). Large grey and white dots indicate dressed
        quantities; solid and curly lines represent quark and gluon propagators,
        respectively. There is a separate quark DSE for the up, down and for
        the strange quarks. The large shaded dot denotes the quenched gluon
        propagator that is taken from the lattice while the quark loop is
        evaluated explicitly. The latter contains an implicit flavour sum over
        up, down and strange.
    }%
\end{figure}

\begin{figure}[b]
    \centering
    \includegraphics[scale=1.0]{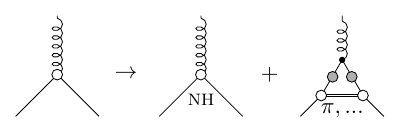}
    \caption{\label{fig:vertex-ansatz}%
        Vertex structure split into non-hadronic parts (NH) and parts including
        potential long-range correlations with quantum numbers of (pseudo)scalar
        mesons.
    }%
\end{figure}

The quark propagator satisfies its associated Dyson--Schwinger equation (DSE),
\begin{equation}
    \label{eq:quark_dse}
    S_{f}^{-1}(p)
    =
    Z_{2}
    \bigl(
        \ii \gamma_{4} \tilde{\omega}_{n}^{f}
        +
        \ii \vec{\gamma} \cdot \vec{p}
        +
        Z_{m} m_{f}
    \bigr)
    -
    \Sigma_{f}(p)
    \,.
\end{equation}
with renormalized bare-quark mass $m_{f}$ and wave-function and mass
renormalization constants $Z_{2}$ and $Z_{m}$. These are calculated in vacuum
using a momentum-subtraction scheme. The quark self energy is given by
\begin{multline}\label{eq:quark_self-energy}
    \Sigma_{f}(p)
    =
    (\ii \gs)^{2}
    \frac{4}{3}
    \frac{Z_{2}}{\tilde{Z}_{3}}
    T
    \sum_{\omega_{n}}
    \int \frac{\dd^{3} q}{(2 \pi)^{3}}
    D_{\nu\rho}(k)
    \gamma_{\nu}
    \times
    \\
    \times
    S_{f}(q)
    \Gamma_{\rho}^{f}(p, q; k)
    \,.
\end{multline}
Here, $k = p - q$ indicates the gluon momentum, $g$ labels the strong coupling
constant, $\tilde{Z}_{3}$ represents the ghost renormalization constant and
$D_{\nu\rho}$ is the dressed gluon propagator. The prefactor of $4/3$ originates
from the colour trace. 
\begin{table*}
	\centering
	\def\arraystretch{1.5}
	\begin{tabular}{|c|c||c|c|c|c|c|}
		\hline
		$m_\pi$ [MeV]         &                      & $0$   &
		$55$                  & $80$                  & $110$
		&
		$140$\\
		\hline\hline
		\multirow{6}{*}{$\Tc \ [\SI{}{MeV}]$} 
		& DSE (this work) &&
		$145.7$               & $147.5$               & $150.2$               &
		$153.3$ \\
		\cline{2-7}
		& DSE ~\cite{Bernhardt:2023hpr}  &$146.7$&
		$149.9$               & $151.6$               & $154.0$               &
		$156.7$ \\
		\cline{2-7}
		& FRG~\cite{Braun:2020ada}                & $142$ &
		$148.0$               & $150.5$               & $153.6$               &
		$156.3$                \\
		\cline{2-7}
		& FRG--DSE~\cite{Gao:2021vsf}                & $141.3$ &
		$146.5$               & $149.1$               & $152.1$               &
		$155.4$                \\
		\cline{2-7}
		& HotQCD ($N_\tau = 12$)~\cite{HotQCD:2019xnw}& -     &
		-                    & $149.7^{+0.3}_{-0.3}$ & $155.6^{+0.6}_{-0.6}$ &
		$158.2^{+0.5}_{-0.5}$  \\
		\cline{2-7}
		& HotQCD ($N_\tau = 8$)~\cite{HotQCD:2019xnw} & -     &
		$150.9^{+0.4}_{-0.4}$ & $153.9^{+0.3}_{-0.3}$ & $157.9^{+0.3}_{-0.3}$ &
		$161.0^{+0.1}_{-0.1}$  \\
		\hline
	\end{tabular}
	\caption{\label{tab:tc-ontheway}%
		Comparison of critical temperatures for different up/down-quark masses
		corresponding to different pion masses and fixed physical strange-quark
		masses between our DSE findings, the FRG, FRG--DSE and the lattice
		results, respectively.%
	}
\end{table*}

A graphical version of the quark DSE is displayed in the top row of
Fig.~\ref{fig:dses}. The gluon propagator is calculated using a simplified
version of the full gluon DSE, illustrated in the bottom row and denoted by:
\begin{equation}\label{eq:gluon_dse}
    D_{\nu\rho}^{-1}(k)
    =
    \bigl[
    D_{\nu\rho}^{\textrm{YM}}(k)
    \bigr]^{-1}
    +
    \Pi_{\nu\rho}(k)
    \,.
\end{equation}
Here, $D_{\nu\rho}^{\textrm{YM}}$ denotes the quenched gluon propagator given by
a combination of all pure Yang--Mills for which we use temperature-dependent
fits to results of quenched lattice calculations
\cite{Fischer:2010fx,Maas:2011ez,Eichmann:2015kfa} as input. The quark part of
the gluon self-energy, $\Pi_{\nu\rho}$, contains a quark loop for every quark
flavour and is calculated explicitly within our framework, see
\cite{Gunkel:2021oya,Bernhardt:2023hpr} for technical details. The last
remaining correlation function in Eq.~(\ref{eq:quark_self-energy}), the dressed
quark--gluon vertex, can be split into non-hadronic parts and parts that involve
meson exchange diagrams visualized in Fig.~\ref{fig:vertex-ansatz}. The latter
contributions include those from (pseudo)scalar mesons that are expected to
become long-ranged at and in the vicinity of second-order phase transitions. In
previous works, we compared results using two different truncations of the
vertex: (i) without taking meson effects into account, and (ii) taking
additionally into account the effects of (off-shell) mesons in two different
approximations, namely (iia) using only pions and sigma mesons
\cite{Gunkel:2021oya} but with chemical-potential-dependent Bethe--Salpeter wave
functions determined in Ref.~\cite{Gunkel:2019xnh,Gunkel:2020wcl} and (iib)
including the full $\SU(3)$ multiplets but with approximated Bethe--Salpeter
wave functions that are only reliable at small (real or imaginary) chemical
potential \cite{Bernhardt:2023hpr} and therefore not suitable for the purpose of
this work. In general, it turned out that while the meson degrees of freedom
control the critical properties of second-order transitions (such as critical
exponents), they only moderately affect the location of the QCD critical
endpoint: In \cite{Gunkel:2021oya}, the CEP has been found at $(\TCEP,\muCEP) =
(117,600) \, \si{\MeV}$ including these fluctuations and at $(\TCEP,\muCEP) =
(112,636) \, \si{\MeV}$ MeV without. Since version (iia) truncations are
computationally much more expensive than version (i) truncations, we will
restrict ourselves to the latter ones in this work. For details, we refer the
reader to \cite{Gunkel:2020wcl}; we use the truncation corresponding to the
first line of Table I in \cite{Gunkel:2020wcl}. As a consequence, we will be
able to gain qualitative information on the changes of the CEP under variation
of $\massl$ with moderate effort. The implied loss of quantitative information
on the level of five percent is a prize we have to pay.

Finally, note that for simplicity we also work with fixed strange chemical potential $\mus=0$. 

\begin{figure*}
	\centering
	\def\arraystretch{1.5}
	\raisebox{60pt}{
		\begin{tabular}{|c||c|c|c|c|}
			\hline
			$m_\pi$ [MeV]   					&	$55$    & $80$    & $110$	& $140$ \\\hline
			\hline
			$T^{\text{CEP}}$ [MeV] 				&   $115$   & $115$   & $113$   & $112$ \\
			\hline
			$\muB^{\text{CEP}}$ [MeV] 			&   $534$  & $552$    & $594$   & $630$ \\
			\hline
			$\kappa_2$    			        	&   $0.0157$  & $0.0157$    & $0.0158$   & $0.0160$ \\
			\hline
	\end{tabular}}
	\hfil
	\includegraphics[scale=0.3]{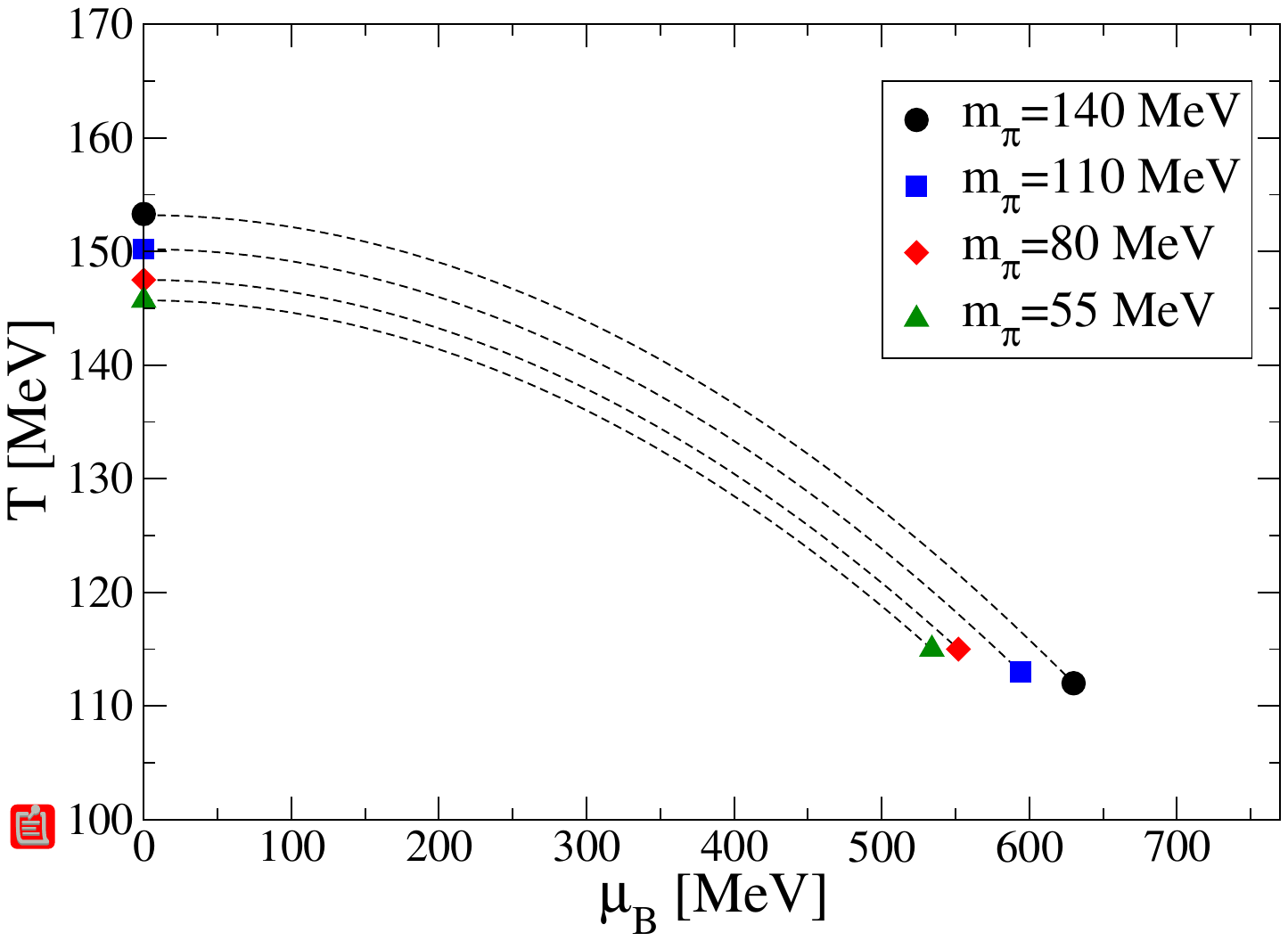}
	\caption{\label{fig:CEP_mass}%
		Left: Data points for the location of the critical endpoint and the 
		curvatures of pseudocritical crossover lines at four different pion 
		masses. Right: Corresponding phase diagram.
	}
\end{figure*}

\section{\label{sec:results}%
    Results and discussion
}

To start with and to gauge our results, we discuss the pseudocritical transition
temperatures for the crossover at zero chemical potential, i.e., along the green
line between the physical point and the left-hand side of the 2$d$-Columbia plot
in Fig.~\ref{fig:columbia-plot}. In table \ref{tab:tc-ontheway}, we display our
results together with previous ones in our framework (including the meson
diagrams; see previous section) \cite{Bernhardt:2023hpr}, results from other
functional approaches \cite{Braun:2020ada,Gao:2021vsf} and from lattice QCD
\cite{HotQCD:2019xnw}. Clearly, the general decreasing trend of the transition
temperatures towards the chiral limit is captured by our approach. Compared to
the much more elaborate calculation of \cite{Bernhardt:2023hpr} our results are
off by only a few MeV.

In Fig.~\ref{fig:CEP_mass} we display our data points for the location of the
critical endpoint at four different pion masses. We clearly see that the
transition temperatures (slowly) rise towards the chiral limit while the
corresponding values of the critical baryon chemical potential decrease. This
behaviour is in agreement with the general trend seen in model calculations
\cite{Halasz:1998qr} and also supports the general notion of temperature bounds
\cite{Karsch:2019mbv} implicit in Fig.~\ref{fig:phases} and discussed in the
introduction. The changes, however, are small and therefore indicate a flat
second-order surface in the 3$d$-Columbia plot, Fig.~\ref{fig:columbia-plot}. A
linear extrapolation of our data points towards the chiral limit indicates the
presence of a tricritical point at $(T^{\text{tri}},\mu_B^{\text{tri}}) =
(117,473)$ MeV, but this result should be taken with more than one grain of
salt. First, there is no convincing reason to assume that a linear extrapolation
should be reliable close to the chiral limit. Second, it may very well be that
the currently omitted mesonic long-range degrees of freedom become more
important in the chiral limit. Third, a direct calculation in the chiral limit
in the truncation scheme used in this work did not confirm this point. Instead,
we encountered numerical problems (absent for the results presented in
Fig.~\ref{fig:CEP_mass}) that prohibited us to extract a trustworthy result. We
have therefore postponed the study of the second-order critical surface close to
the chiral limit to a future work conducted in a type (iia) truncation scheme as
discussed in the previous section.

From the results presented in Fig.~\ref{fig:CEP_mass}, we are able to
analytically extract the corresponding curvature coefficient $\kappa_2$ of the
pseudocritical transition line from the approximate expression
\begin{equation}
    T(\muB) 
    = 
    T(\muB=0) \left(
        1 - \kappa_2 \left(
            \frac{\muB}{T(\muB=0)}
        \right)^2
    \right)
\end{equation}
assuming all higher-order coefficients to be zero. While the value at physical
light-quark masses agrees well with the high-quality extractions from functional
approaches \cite{Fu:2019hdw,Gao:2020fbl,Gunkel:2021oya} and lattice QCD
\cite{Bonati:2018nut,HotQCD:2018pds,Borsanyi:2020fev}, it is interesting to note
that the curvature is more or less constant with varying light-quark masses and
only very slightly decreases towards the chiral limit. This trend seems to be
somewhat in contrast to the one extracted in a very elaborate scaling analysis
performed in Ref.~\cite{Ding:2024sux}. For several reasons, however, this should
not be overemphasized. First, our numbers do not originate from an expansion
around $\muB=0$. They are extracted assuming a very simple expression to hold
over a large range of chemical potential, which is probably not justified.
Second, our numbers do not originate from a scaling analysis. In order to
compare apples with apples, we would need to use our more advanced truncation
schemes explicitly including the potentially long-ranged meson degrees of
freedom, Refs.~\cite{Gunkel:2021oya,Bernhardt:2023hpr}, and perform a complete
scaling analysis around $\muB=0$ taking care also of variations of the
strange-quark chemical potential. This is outside the scope of the present work.

Finally, we also checked the order of the phase transition at imaginary chemical
potential up to values approaching the Roberge--Weiss transition. At all four
values of the pion mass, we choose values of imaginary baryon chemical potential
along lines of constant $\muB/(\pi T) \in \{0, 1/8, 3/8, 5/8, 7/8\}$ (with $\muB
= 3 \mu_{\ell}$ in our case) and evaluate the subtracted condensate along those
lines. The corresponding susceptibility is smooth, indicating a crossover in
each case in accordance with previous indications from lattice gauge theory
\cite{DAmbrosio:2022kig,Cuteri:2022vwk}.

\section{\label{sec:summary}%
    Summary and conclusions
}

In this work, we performed a qualitative study of the nature of the chiral
transition of QCD at fixed and physical values of the strange-quark mass and
variations of the degenerate light-quark masses between the physical point and
the chiral limit. To this end, we employed a coupled set of Dyson--Schwinger
equations in a simplified truncation scheme that delivers results for the
location of the critical endpoint close to our currently most-advanced scheme
\cite{Gunkel:2021oya}. We identify a flat chiral critical surface at large
chemical potential that presumably ends in a line of tricritical points along
the chiral left-hand side of the 3$d$-Columbia plot, see
Fig.~\ref{fig:columbia-plot}. Inside the green area in this plot, we furthermore
find crossover transitions throughout indicating a large volume of crossovers in
the 3$d$-Columbia plot. Our findings are in agreement with and confirm previous
notions from model approaches and lattice gauge theory concerning temperature
bounds for the transition temperature of the critical endpoint
\cite{Halasz:1998qr,Karsch:2019mbv} and the structure of the Columbia plot
\cite{deForcrand:2017cgb,HotQCD:2019xnw,Cuteri:2021ikv,Cuteri:2021ikv,Dini:2021hug,DAmbrosio:2022kig,Cuteri:2022vwk}. 
Our results are in agreement with the qualitative behaviour sketched in 
Fig.~\ref{fig:phases} and the corresponding existence of a tricritical point 
at finite chemical potential in the light-quark chiral limit of QCD. 
However, we did not succeed in identifying the location of this point. 
This remains for future work.


\section{Acknowledgments}
We thank Jens Braun, Frithjof Karsch, Owe Philipsen, Fabian Rennecke,
Bernd-Jochen Schaefer and Lorenz von Smekal for fruitful discussions. This work
has been supported by the Helmholtz Graduate School for Hadron and Ion Research
(HGS-HIRe) for FAIR, the GSI Helm\-holtz\-zentrum f\"{u}r Schwerionenforschung, and
the Deut\-sche Forschungsgemeinschaft (DFG, German Research Foundation) through
the Collaborative Research Center TransRegio CRC-TR 211 ``Strong-interaction
matter under extreme conditions'' and the individual grant FI 970/16-1.


\bibliographystyle{utphys_mod}
\bibliography{ColumbiaMesonBCBibliography}

\end{document}